\documentclass[11pt]{article}
\usepackage[utf8]{inputenc}	
\usepackage{amsmath,amsthm,amsfonts,amssymb,amscd}
\usepackage{multirow,booktabs}
\usepackage[table]{xcolor}
\usepackage[a4paper,margin=1in]{geometry}
\usepackage{lastpage}
\usepackage{enumitem}
\usepackage{fancyhdr}
\usepackage{mathrsfs}
\usepackage{wrapfig}
\usepackage{setspace}
\usepackage{subcaption}
\usepackage{graphicx}
\usepackage{calc}
\usepackage{multicol}
\usepackage{cancel}
\usepackage[most]{tcolorbox}
\usepackage{xcolor}
\usepackage{hyperref}
\usepackage{tikz}
\usetikzlibrary{external}
\usepackage{caption}
\usepackage{accents}
\usepackage{float}

\usepackage[maxbibnames=99, style=numeric]{biblatex}
\pdfoutput=1

\newtheorem{theorem}{Theorem}[section]

\newtheorem{definition}[theorem]{Definition}

\newtheorem{proposition}[theorem]{Proposition}

\theoremstyle{remark}
\newtheorem{remark}[theorem]{Remark}
\newtheorem{example}[theorem]{Example}

\DeclareMathOperator{\supp}{supp}

\DeclareMathOperator{\id}{Id}
 \DeclareMathOperator{\mean}{mean}

\newcommand{\Acal}{\mathcal{A}}
\newcommand{\Bcal}{\mathcal{B}}

\newcommand{\Fcal}{\mathcal{F}}

\newcommand{\Pcal}{\mathcal{P}}
\newcommand{\Qcal}{\mathcal{Q}}

\newcommand{\Wcal}{\mathcal{W}}

\newcommand{\EE}{\mathbb{E}}

\newcommand{\NN}{\mathbb{N}}

\newcommand{\PP}{\mathbb{P}}

\newcommand{\RR}{\mathbb{R}}

\usepackage{bbm}

\renewcommand{\epsilon}{\varepsilon}

\newcommand{\Law}{\mathscr L}

\newcommand{\cadlag}{c\`{a}dl\`{a}g}

\title{Dynamic reinsurance via martingale transport} 

\author{Beatrice Acciaio\thanks{Department of Mathematics, ETH Z\"{u}rich, Switzerland. \texttt{beatrice.acciaio@math.ethz.ch}}, Brandon Garcia Flores\thanks{Centrica Trading A/S, Denmark.}, Antonio Marini\thanks{Department of Mathematics, ETH Z\"{u}rich, Switzerland. \texttt{antonio.marini@math.ethz.ch}}, and Gudmund Pammer\thanks{Institut für Statistik	, TU Graz, Austria. \texttt{gudmund.pammer@tugraz.at}}}
\date{\today }

\begin{document}

\maketitle

\begin{abstract} We formulate a dynamic reinsurance problem in which the insurer seeks to satisfy prescribed terminal moment or risk-based constraints while minimizing the $L^2$-norm of the ceded risk. As a tool for this analysis, we first use techniques from martingale optimal transport to study the auxiliary problem in which the insurer matches a given terminal distribution of the surplus process. We show that, under suitable assumptions, this auxiliary problem admits a tractable solution analogous to the Bass martingale. We then relax this condition by only requiring certain moment or risk-based constraints.
\\

\end{abstract}

\section{Introduction}\label{sect:intro}
A central problem in mathematical insurance is the design of optimal reinsurance strategies, motivated by the need to reduce excessive risk retained by the primary insurer, often in response to regulatory requirements. These considerations have led to the study of the optimal reinsurance problem, which has been extensively studied in the literature. Classical references include those of de Finetti \cite{definetti1940}, Borch \cite{borch1960attempt,borch1960safety}, and Arrow \cite{arrow63}, where the problem is studied in a variety of settings. The topic is still a very active field of research, see for instance \cite{centeno} and \cite{albrecher2017reinsurance} for an overview.

In this paper, we will focus on  \textit{dynamic reinsurance}. In contrast to the classical static approach, the idea is to relax the assumption that the contract is fixed at the moment of inception and instead allow adaptations  along the duration of the contract. We will assume that these adaptations can be done in a continuous-time setup. 
A large portion of the existing literature on dynamic reinsurance has been motivated by classical risk measures such as the infinite-horizon probability of ruin and the finite-time mean–variance criteria, which are usually considered common benchmarks. Alongside these performance metrics, the literature has traditionally concentrated on specific contract forms, such as fixed quota-share (QS) or fixed excess-of-loss (XL) treaties, and natural generalizations including multilayered contracts; see \cite{schmidli2001optimal,hipp2001optimal, zhou2014optimal}. Extensions of the problem combine the search for optimal reinsurance with dividend optimization \cite{azcue2005optimal}, capital injection minimization \cite{eisenberg2009optimal, eisenberg2011minimising}, and utility maximization for the insurer and reinsurer \cite{korn2012worst}, often combined with the use of an approximation of the risk process by a Brownian motion. Another line of research integrates financial considerations into the problem, by allowing the insurance company not only to manage reinsurance but also to invest premium income in financial markets, typically including both risk-free and risky assets; see \cite{hipp2000optimal, taksar2003optimal, yang2005optimal, liu2004optimal}. From a methodological point of view, the most common approach to solving these optimization problems is to recast them into a dynamic programming framework and study the associated Hamilton–Jacobi–Bellman (HJB) equations.

In this paper, we study an optimal reinsurance problem in which an insurance company seeks to purchase reinsurance so as to satisfy prescribed terminal constraints, such as moment or risk-based constraints, on the distribution of its surplus at a fixed time horizon, while minimizing the $L^2$-norm of the ceded risk. To address this problem, we use optimal transport as a tool: we first study the auxiliary problem in which the terminal surplus distribution is prescribed, and then exploit this formulation to handle the constrained reinsurance problem.

Since the seminal works of Monge \cite{Monge} and Kantorovich \cite{Kant42}, optimal transport has developed into a powerful and versatile tool with applications across numerous fields, including logistics, engineering, data science, and finance. Martingale optimal transport is a variant of optimal transport in which the admissible transport plans are required to satisfy a martingale constraint. Originally introduced in robust finance, it has been used both to compute model-independent bounds for derivative prices and to select models matching prescribed market marginals while remaining close to a reference dynamics; see \cite{HoNe12, BeHePe12, HeTaTo16, DoSo12, AcBePeSc16, He17, GuLoWa19, NuWiZh22}.
A central development in this area is the Martingale Benamou--Brenier problem \cite{BaBeHuKa20}, where one looks for a martingale with prescribed marginals that remains as close as possible to a reference martingale. This point of view has led to explicit and tractable constructions, such as the Bass martingale and its variants, which can be efficiently simulated; see \cite{CoHe21, AcMaPa23, JoLoOb24, Ts24}. Our approach is inspired by this perspective, though the setting considered here is simpler: the martingales we optimize over start from deterministic initial values. As a consequence, the main tool required in our analysis is the classical quadratic Monge--Kantorovich problem at the terminal time, rather than the full machinery of martingale optimal transport.

In the present work, we adopt as cost functional for our reinsurance problem the quadratic variation of the difference between the surplus processes before and after reinsurance. We show that this formulation admits a tractable optimizer that closely mimics the dynamics of the original surplus process, analogous to the way in which the Bass martingale replicates the dynamics of Brownian motion. 
 Under suitable assumptions, this cost coincides with the $L^2$-norm of the risk ceded in time. Specifically, we assume that any admissible surplus process can be decomposed into the sum of a deterministic function and a pure-jump martingale. A classic and widely studied example satisfying this structure is the Cramér--Lundberg model. Another application of optimal transport to reinsurance appears in \cite{acciaio2025optimal}, although the perspective adopted there is substantially different. 
Indeed, \cite{acciaio2025optimal} considers a static perspective, looking for a reinsurance contract such that the terminal retained exposure satisfies certain constraints, and the objective to be minimized is some functional of the ceded risk.
In contrast, the present paper adopts the dynamic surplus-process perspective described above.
Nevertheless, we will be able to reformulate the dynamic transport problem as a static one, so that we can use classical results to find the optimal reinsurance contract. 

Finally, we extend our framework to a relaxed setting, where the distribution of the surplus process at time $T$ is not fixed, but instead required to satisfy certain moment or risk-based constraints, such as variance, (average) value-at-risk, skewness, or kurtosis. This leads to a double optimization problem which, under suitable assumptions, can be reduced to a convex optimization problem by leveraging properties of the Bass martingale. This extension demonstrates that the proposed approach remains effective even when the insurer does not aim to match an exact terminal distribution.

\paragraph{Notations.}
\begin{itemize}
\item We write $\Pcal(\RR)$ for the probability measures on $\RR$ and $\Pcal_p(\RR)$ for the subset of probability measures with finite $p$-moment, $p\in[1,\infty)$.
\item For any $\xi\in\Pcal(\RR)$, we use $F_\xi$ and $Q_\xi$ to denote the corresponding CDF and quantile function. More precisely, $F_\xi(x):=\xi((-\infty,x])$ and $Q_\xi(u):=\inf{x\in\RR:F_\xi(x)>u}$, $u\in(0,1)$, so that $Q_\xi$ is right-continuous. Whenever it is finite, we set $\operatorname{mean}(\xi):=\int_\RR x,\xi(dx)$. We also denote by $\supp(\xi)$ the support of $\xi$ and by $\delta_x$ the Dirac measure at $x$.
\item For a random variable $X$, we denote its law by $\Law(X)$ and write $X\sim\xi$ when $\Law(X)=\xi$. We use the abbreviations $F_X:=F_{\Law(X)}$ and $Q_X:=Q_{\Law(X)}$.
\item We denote by $\gamma$ the standard normal distribution, with density $\phi$ and CDF $\Phi$, and we write $\lambda$ for the Lebesgue measure on $\mathbb{R}$.
\item For $\mu,\nu\in \mathcal{P}(\mathbb{R})$, we denote by $\Pi(\mu,\nu)$ the subset of $\mathcal{P}(\mathbb{R}^2)$ consisting of measures with first marginal $\mu$ and second marginal $\nu$; its elements are called couplings of $\mu$ and $\nu$.
We write $\mathcal{M}(\mu,\nu)\subset \Pi(\mu,\nu)$ for the set of $\pi$ such that $\operatorname{mean}(\pi^x)=x$ for $\mu$-a.e. $x$, where $\pi(dx,dy)=\mu(dx)\pi^x(dy)$ is the disintegration of $\pi$ with respect to $\mu$. The elements of $\mathcal{M}(\mu,\nu)$ are the martingale couplings of $\mu$ and $\nu$.
\item The push-forward measure of $\xi \in \Pcal(\RR)$ through a measurable map $T:\RR\rightarrow\RR$, denoted by $T_\#\xi$, is the probability measure such that $T_\#\xi(A)=\xi(T^{-1}(A))$, for any $A\in\Bcal(\RR)$. We denote by $\mathrm{Lip}_1(\RR)$ the set of $1$-Lipschitz functions from $\RR$ to itself.
\item For $\mu,\nu\in\Pcal_1(\RR)$, we say that $\mu$ is dominated in convex order by $\nu$, and write $\mu\preceq_c\nu$, if for all convex functions $h:\RR\to\RR$ with linear growth we have $\int h,d\mu\leq\int h,d\nu$.
\item For $\xi,\zeta\in\Pcal(\RR)$, we write $\xi\ast\zeta$ for the probability measure representing their convolution, so that $\xi\ast\zeta(A)=\int \mathbf{1}_A(x+y),\xi(dx)\zeta(dy)$, $A\in\Bcal(\RR)$. For two measurable functions $f,g$, their convolution is the function given by $(f\ast g)(x)=\int f(x-y)g(y),dy$, $x\in\RR$. Moreover, the convolution of $f$ and $\xi$ is the function defined as $(f\ast\xi)(x)=\int f(x-y),\xi(dy)$, $x\in\RR$. We further define $(\xi\star f)(x):=\int f(x+y),\xi(dy)$, $x\in\RR$, whenever the integral is well defined.
\end{itemize}

\section{Background on (Martingale) Optimal Transport Theory} \label{sect:OT}
In this section we recall the  optimal transport problem and its martingale variant, together with the main tools of the transport theory that will be used in the paper.

Optimal transport theory originates from a problem posed by Gaspard Monge in 1781 \cite{Monge}, concerning the most efficient way to transport one (probability) distribution $\mu$ into another $\nu$. In its modern formulation, due to Kantorovich \cite{Kant42}, and for quadratic cost on $\RR^d$, optimal transport leads to the $2$-Wasserstein distance $\Wcal_2$, defined by
\begin{equation}
    \label{MK}
    \Wcal_2^2(\mu,\nu) := \inf_{\pi \in \Pi(\mu,\nu)} \int_{\RR^d \times \RR^d} |x-y|^2\,d\pi(x,y),
\end{equation}
where $\Pi(\mu,\nu)$ denotes the set of probability measures on $\RR^d \times \RR^d$ with marginals $\mu$ and $\nu$. Its elements are called transport plans, and the minimizers in \eqref{MK} are called optimal transport plans.
Equivalently, in probabilistic language, one may write
\[
    \Wcal_2^2(\mu,\nu) = \inf_{X \sim \mu,\, Y \sim \nu} \EE\bigl[|X-Y|^2\bigr].
\]
When an optimal transport plan is induced by a measurable map $T:\RR^d \to \RR^d$, namely
\[
    \pi = (\mathrm{id},T)_\#\mu, \qquad T_\#\mu=\nu,
\]
the map $T$ is called an optimal transport map. In this case,
\[
    \Wcal_2^2(\mu,\nu) = \int_{\RR^d} |x-T(x)|^2\,d\mu(x).
\]
Thus, an optimal transport map, when it exists, provides a deterministic realization of an optimal transport plan. A fundamental result ensuring the existence and uniqueness of such a map under suitable assumptions is Brenier's Theorem, which we recall here since it will be invoked at several points in this paper.

\begin{theorem}[Brenier's Theorem, \cite{Br87}]\label{thm.brenier}
Let $\mu,\nu \in \mathcal P_2(\RR^d)$, with $\mu \ll \lambda$, and let $c(x,y)=|x-y|^2$.
Then there exists a unique optimal transport plan $\pi \in \Pi(\mu,\nu)$, given by
\[
    \pi = (\mathrm{id},\nabla \varphi)_\#\mu,
\]
for some convex function $\varphi:\RR^d \to \RR \cup \{+\infty\}$. In particular, $\nabla\varphi$ is an optimal transport map from $\mu$ to $\nu$.
Additionally, if $T:\RR^d \to \RR^d$ is a transport map from $\mu$ to $\nu$ and $T=\nabla \tilde\varphi$ for some convex function $\tilde\varphi$, then $T$ is the unique optimal transport map.
\end{theorem}

The one-dimensional case plays a particularly important role and admits a more explicit description. If $\mu,\nu \in \mathcal P_2(\RR)$, the optimal coupling is given by the monotone rearrangement: for $U \sim \mathrm{Unif}(0,1)$, the pair
\[
    \big(Q_\mu(U),Q_\nu(U)\big)
\]
is an optimal coupling between $\mu$ and $\nu$, where $Q_\mu$ and $Q_\nu$ denote the corresponding quantile functions. Equivalently, when the transport map is well defined, for instance when $\mu$ is absolutely continuous, the optimal map from $\mu$ to $\nu$ is given by
\[
    T = Q_\nu \circ F_\mu,
\]
where $F_\mu$ is the cumulative distribution function of $\mu$. In particular, in dimension one the $2$-Wasserstein distance has the simple representation
\[
    \Wcal_2^2(\mu,\nu) = \int_0^1 |Q_\mu(u)-Q_\nu(u)|^2\,du,
\]
so that $\Wcal_2(\mu,\nu)$ is precisely the $L^2$-distance between the quantile functions of $\mu$ and $\nu$.

The modern formulation of optimal transport theory has been significantly influenced by seminal works such as \cite{Br87, BB00, Mc95, JoKiOt98}. Besides the static Monge--Kantorovich formulation recalled above, optimal transport also admits dynamic formulations, which describe the transport between two marginals as the evolution of a stochastic process or of a curve of probability measures.
In recent years, this point of view has been developed in mathematical finance through martingale optimal transport. In a static framework, one imposes on the transport plan the additional constraint of martingality, leading to optimization problems over $1$-step martingales with prescribed marginal distributions.
A dynamic counterpart of this theory is the Martingale Benamou--Brenier problem introduced in \cite{BaBeHuKa20}. In its one-dimensional form, it can be written as
\[
    \inf_{\substack{
    M_t = M_0+\int_0^t \sigma_s\,dB_s \\
    M_0\sim\mu,\; M_1\sim\nu
    }}
    \EE\left[
    \int_0^1 |\sigma_t-1|^2\,dt
    \right],
\]
where the continuous-time martingale $M=(M_t)_{t\in[0,1]}$ is required to have prescribed initial and terminal marginals. Thus, the problem looks for the Brownian martingale with marginals $\mu$ and $\nu$ whose volatility stays as close as possible to the unit volatility of Brownian motion. Its optimizer is known as the stretched Brownian motion, or Bass martingale, and can be viewed as a randomized version of the original Bass martingale construction of \cite{Ba83}. In the case of a deterministic initial marginal, it reduces to the classical Bass form
\[
    M_t
    =
    \EE\big[\varphi'(B_1)\mid B_t\big],
    \qquad t\in[0,1],
\]
for a suitable convex function $\varphi$ chosen so that the terminal marginal constraint is satisfied.

This construction was further generalized in \cite{Ts24} through the notion of $q$-Bass martingales, where the Brownian reference is replaced by a more general reference law or reference process. These results motivate the use of optimal transport techniques to construct martingales with prescribed terminal behavior. In the reinsurance setting, this perspective requires a continuous-time adaptation, since \cite{Ts24} considers discrete-time martingales. In the next section, we implement this adaptation through the same conditional-expectation mechanism, using a different process with independent increments associated with the initial reinsurance strategy (see Subsection \ref{sec:OT_adaptation}). The corresponding framework and the reduction to a classical quadratic optimal transport problem will be introduced there.

\section{Reinsurance strategies and the auxiliary prescribed-law problem}\label{sect:applications}

The aim of this section is twofold. We first introduce the surplus process and the class of admissible reinsurance strategies. We then study an auxiliary problem in which the terminal law of the surplus after reinsurance is prescribed. This prescribed-law problem will serve as the optimal-transport tool used in the next section to address the actual constrained reinsurance problem, where the terminal distribution is only required to satisfy moment or risk-based constraints.

\subsection{Surplus processes and admissible reinsurance strategies}

To formulate the reinsurance problem in a general setting, we begin by introducing the surplus process $(U_t)_{t \geq 0}$ of the insurance company under minimal structural assumptions. Let $(\Omega, \Fcal, \PP)$ be a probability space. We assume that $U_t$ takes the form 
\begin{equation}
\label{general}
	U_t = u_0 + g(t) + M_t,\quad t\geq 0,
\end{equation}
where $u_0 \geq 0$ is the initial capital, $g:[0, \infty) \rightarrow \RR$ is a deterministic function with $g(0)=0$, and $M$ is a martingale w.r.t. its natural filtration $\Fcal^M$, satisfying $M_0 = 0$ a.s. Additionally, we assume that $M$ has the form
\begin{equation*}
    M_t = \EE[S_t] - S_t,\quad t\geq 0,
\end{equation*}
where $S=(S_t)_{t\geq 0}$ is a non-decreasing, pure-jump process with independent increments, representing the insurer's original cumulative claims paid over the interval $[0,t]$. Therefore, the deterministic function $g(t)$ represents the gain generated up to time $t$ in excess of expected claim payments.

\begin{example}
\label{ex:cramer-lundberg}
The Cramér--Lundberg model satisfies the above assumptions. Indeed, in this model the surplus process is given by
\begin{equation}
\label{CL-model}
	U_t = u_0 + (1+\theta)\overline \xi \lambda t - \sum_{i=1}^{N_t} \xi_i,\quad t\geq 0,
\end{equation}
where $\theta>0$ is the safety loading chosen by the insurer, $N=(N_t)_{t\geq 0}$ is a Poisson process with intensity $\lambda > 0$, and $(\xi_i)_{i \in \NN}$ is a sequence of positive i.i.d. random variables, with expected value $\overline \xi$, representing the claims. From this, it is immediate to see that the model admits the decomposition \eqref{general}, with $g(t) = \theta \overline \xi \lambda t$ and $S_t = \sum_{i=1}^{N_t} \xi_i$.
\end{example}

In the reinsurance problem for surplus processes as in \eqref{general}, we assume that the insurer has full flexibility in determining the portion of each claim to retain. Since $S$ is the cumulative-claim process, whenever $\Delta S_t>0$ the jump $\Delta S_t$ represents the claim occurring at time $t$. We describe a reinsurance strategy through the cumulative retained-claim process $S^R$: at each such jump time, the insurer retains the amount $\Delta S_t^R$ and cedes the remaining part $\Delta S_t-\Delta S_t^R$ to the reinsurer. The condition $0\leq \Delta S_t^R\leq \Delta S_t$ ensures that the strategy neither creates additional claims nor retains more than the realized claim amount. The premium paid to the reinsurer corresponds to the compensator of the ceded risk process, possibly augmented by a given deterministic surcharge $h : [0, T] \rightarrow \RR$, which may depend on the law of the surplus process. Thus, after subtracting the compensator of the ceded part from the original surplus, the remaining martingale component is naturally written in terms of the compensator $A^R$ of the retained risk process $S^R$, while the deterministic surcharge reduces the drift from $g$ to $g^R=g-h$. This leads to the following class of admissible reinsurance strategies:

\begin{definition}[Admissible strategy]
A reinsurance strategy is said to be admissible if the surplus process after reinsurance admits the decomposition
\begin{equation}
 \label{UR}
 	U_t^R = u_0^R + g^R(t) + M_t^R,
 \end{equation}
 where
 \begin{equation*}
 	M_t^R = A_t^R - S_t^R, \quad \text{and} \quad g^R(t) = g(t)-h(t).
 \end{equation*}
 Here,
 \begin{itemize}
 	\item $u_0^R \geq 0$ is the initial capital after reinsurance,
 	\item $S^R$ is a càdlàg, $\Fcal^M$-adapted, non-decreasing pure-jump process such that $S_0^R = 0$,
 	\item $A^R$ is the compensator of $S^R$ with respect to the filtration $\Fcal^M$,
 \end{itemize}
and the jump sizes of $S^R$ satisfy
\begin{equation}
	\label{adminissibilty}
	0 \leq \Delta S^R_t \leq \Delta S_t \text{ a.s.}\quad \text{for any } t \in [0, T].\footnote{For any càdlàg process $X$, we set $\Delta X_t := X_t - X_{t-}$.}
\end{equation}

\end{definition}

\begin{remark}
	The process $S^R$ represents the claims retained by the insurer. Since $S^R$ is adapted to $\Fcal^M$, it can only jump when the original claim process $S$ does. The jump size constraint \eqref{adminissibilty} further ensures that the retained portion of each individual claim does not exceed the size of the corresponding original claim.
\end{remark}

The class of admissible reinsurance strategies introduced above is intentionally broad. No monotonicity of the retained and ceded risk with respect to the claim size is imposed a priori, as is usual in formulations based on no-sabotage or moral-hazard restrictions; see, for instance, \cite{Ch12, GhJiRe22}. Our framework includes, in particular, standard dynamic treaty forms whenever the corresponding retention rule is adapted to the available information. For example, proportional reinsurance can be represented by $\Delta S_t^R = q_t \Delta S_t$, while an excess-of-loss retention can be written as $\Delta S_t^R = \min(\Delta S_t, d_t)$, for suitable adapted processes $q_t \in [0,1]$ and $d_t \geq 0$. Thus, this formulation contains familiar treaties, such as those considered in \cite{CeSi91}, but also allows more general state-dependent arrangements. In this broader class, an optimizer need not correspond to a standard marketable treaty and may instead be interpreted as a theoretical benchmark unless additional contractual restrictions are imposed. A similarly broad point of view on admissible reinsurance arrangements is adopted in \cite{acciaio2025optimal}.

\subsection{An optimal-transport tool for prescribed terminal laws}\label{sec:OT_adaptation}

Before applying the construction to reinsurance, we recall the abstract martingale problem that motivates it. Let $(X_t)_{t \in [0,T]}$ be a process with independent and square-integrable increments, and $\Fcal^X$ its canonical filtration. We denote by $\overline X$ the compensator of $X$, so that
\[
	\widetilde X_t := X_t-\overline X_t
\]
is a square-integrable martingale. Given a target distribution $\nu \in \Pcal_2(\RR)$, we are interested in finding a martingale $(N_t)_{t \in [0,T]}$, adapted to the same filtration $\Fcal^X$, with deterministic initial value $N_0=\mean(\nu)$ and terminal distribution $N_T \sim \nu$, which minimizes the expected quadratic variation of the difference $N-X$. More precisely, we consider
\begin{equation}
\label{eq:general_Pb}
    \inf_{\substack{N_0=\mean(\nu),\, N_T \sim \nu\\ (N_t)_{t \in [0,T]} \text{ is a martingale w.r.t. } \Fcal^X}}
    \EE \big[\langle N-X \rangle_T\big].
\end{equation}

The optimization problem in \eqref{eq:general_Pb} is path-dependent, since the quadratic variation of $N-X$ is a functional of the whole trajectory of the process on $[0,T]$. Nevertheless, its optimizer can be identified by solving a classical quadratic optimal transport problem at the terminal time $T$. Indeed, since $\overline X$ has finite variation, it does not contribute to the quadratic variation, and therefore
\[
	\langle N-X\rangle_T = \langle N-\widetilde X\rangle_T .
\]
Moreover, $N-\widetilde X$ is a square-integrable martingale. Hence, using the martingale isometry,
\[
	\EE\big[\langle N-X\rangle_T\big]
	=
	\EE\big[\langle N-\widetilde X\rangle_T\big]
	=
	\EE\big[(N_T-\widetilde X_T)^2\big] - (N_0-\widetilde X_0)^2 .
\]
Since the last term is fixed over the admissible class, the minimization \eqref{eq:general_Pb} is equivalent to the minimization
\[
    \inf_{\substack{N_0=\mean(\nu),\, N_T \sim \nu\\ (N_t)_{t \in [0,T]} \text{ is a martingale w.r.t. } \Fcal^X}}
    \EE \big[(N_T-\widetilde X_T)^2\big].
\]
Thus, up to an additive constant, the dynamic problem reduces to the classical quadratic optimal transport problem between $\Law(\widetilde X_T)$ and $\nu$. In particular, if $X$ is already a martingale, so that $\overline X=0$, this is simply the quadratic optimal transport problem between $\Law(X_T)$ and $\nu$.

Assume now that the optimal transport from $\Law(\widetilde X_T)$ to $\nu$ is induced by a map $F:\RR \to \RR$. In the one-dimensional setting recalled above, this map is the monotone rearrangement; for instance, when $\Law(\widetilde X_T)$ is absolutely continuous, it is given by
\[
	F = Q_\nu \circ F_{\Law(\widetilde X_T)},
\]
where $F_{\Law(\widetilde X_T)}$ denotes the cumulative distribution function of $\widetilde X_T$ and $Q_\nu$ is the quantile function of $\nu$.

The optimizer of \eqref{eq:general_Pb} is then obtained by projecting the optimal terminal value onto the filtration of $X$, namely
\[
	N_t^* := \EE\big[F(\widetilde X_T)\mid \Fcal_t^X\big], \qquad t \in [0,T].
\]
Indeed, by construction, $(N_t^*)_{t \in [0,T]}$ is an $\Fcal^X$-martingale, $N_0^*=\mean(\nu)$, and $N_T^*=F(\widetilde X_T)\sim \nu$. Moreover, the pair $(\widetilde X_T,N_T^*)$ realizes the optimal quadratic coupling between $\Law(\widetilde X_T)$ and $\nu$, and therefore $N^*$ minimizes \eqref{eq:general_Pb}.

The martingale structure also provides a dynamic optimality property. Since $\widetilde X$ has independent increments, the conditional expectation above can be written as
\[
	N_t^* = F_t(\widetilde X_t),
\]
for a deterministic function $F_t$. Moreover, if $F$ is monotone, then $F_t$ is monotone as well. Consequently, for every $t \in [0,T]$, the pair $(\widetilde X_t,N_t^*)$ is again an optimal quadratic coupling between $\Law(\widetilde X_t)$ and $\Law(N_t^*)$. Therefore, the restriction of $N^*$ to any time interval $[0,t]$ solves the analogous problem with terminal target distribution $\Law(N_t^*)$.

This shows that, although the original formulation \eqref{eq:general_Pb} involves a pathwise criterion, its optimizer can be constructed through a one-dimensional classical quadratic optimal transport problem at the terminal time. The martingale property then propagates optimality backward to all intermediate times. This approach is inspired by the Martingale Benamou--Brenier problem \cite{BaBeHuKa20} and by the theory of $q$-Bass martingales \cite{Ts24}. In the present setting, however, the analysis relies only on classical quadratic optimal transport, since the martingales under consideration start from deterministic initial values.

\subsection{The prescribed-law reinsurance problem}

We now apply the previous construction to reinsurance. Our final objective is to design reinsurance strategies for which the terminal surplus satisfies prescribed constraints, such as moment or risk-based constraints, while minimizing the $L^2$-norm of the ceded risk. Before addressing this constrained problem, we introduce an auxiliary problem in which the terminal law of the surplus after reinsurance is fixed. This prescribed-law formulation is not the final object of the paper, but rather the optimal-transport tool that will be used in the next section to treat more flexible terminal specifications.

More precisely, given a target distribution $\nu \in \Pcal_2(\RR)$ and a fixed time horizon $T>0$, we first require the surplus process after reinsurance to satisfy $U_T^R\sim \nu$. Among all admissible reinsurance strategies with this terminal law, we minimize the $L^2$-norm of the ceded risk. This leads to the auxiliary problem

\begin{equation}
 \label{mainPb}
 	\inf_{\substack{U^R \text{ is admissible} \\ U_T^R \sim \nu}}
 	\sum_{t \in [0,T]}
 	\EE \left[ |\Delta S_t^R - \Delta S_t|^2  \right],
\end{equation}
where the cost corresponds to the cumulative quadratic size of the ceded claim amounts.

In the Cramér--Lundberg model (see Example \ref{ex:cramer-lundberg}), the cost functional in \eqref{mainPb} admits a direct risk interpretation. Suppose that the same deterministic ceded-loss function $I$ is applied to each claim, so that the aggregate ceded loss up to time $T$ is $\sum_{i=1}^{N_T} I(\xi_i)$. Its variance is then $\lambda T\,\EE[I(\xi_1)^2]$, which coincides with the quadratic cost in \eqref{mainPb}. Thus, for such a static claim-by-claim treaty, our criterion is exactly the variance of the aggregate ceded loss, providing a direct connection with the second-order risk criteria arising in mean--variance and variance-related reinsurance problems; see, for instance, \cite{Ka01, Ka04}. Related criteria also appear in \cite{Ch12, AsChHu15}, but in different roles within the optimization problem. In \cite{Ch12}, VaR and CVaR are minimized under a variance-related premium principle, while \cite{AsChHu15} considers risk exposure under Solvency II, incorporating retained risk, reinsurance premiums, and risk-margin effects. Here, instead, the quadratic criterion measures the amount of risk transferred to the reinsurer, while the insurer's terminal risk profile is imposed through constraints. In the prescribed-law problem \eqref{mainPb}, this profile is specified by fixing the terminal distribution of the insurer's surplus.

The solution of \eqref{mainPb} provides the basic building block for the constrained reinsurance problem studied in the next section.

\subsection{Optimal transport solution of the auxiliary problem}

In this section, we investigate the solution to problem \eqref{mainPb} and identify the conditions ensuring that the Bass construction provides an optimal reinsurance strategy.
This yields an optimal reinsurance strategy that is Markovian.

\begin{theorem}
\label{mainThm}
Assume there exists an increasing $1$-Lipschitz map $F:\RR \rightarrow \RR$ such that $F(M_T) \sim \nu$. Then the optimizer of \eqref{mainPb} is given by
\begin{equation*}
    \hat U_t^R = \hat u_0^R + g^R(t) + \hat M_t^R,
\end{equation*}
where 
\begin{equation*}
	\hat M_t^R = \EE[\hat F(M_T)|M_t], \quad \hat F = F - \text{mean}(\nu), \quad \text{and} \quad \hat u_0^R=\text{mean}(\nu)-g^R(T).
\end{equation*}
\end{theorem}
The requirement for the map $F$ to be $1$-Lipschitz is crucial as it ensures admissibility of the strategy.
Indeed, monotonicity ensures that the retained-claim process $S^R$ is a \cadlag, $\Fcal^M$-adapted, non-decreasing pure-jump process, while the $1$-Lipschitz property ensures that its jump sizes satisfy \eqref{adminissibilty}.
This condition also has a weak-transport interpretation; see \cite{GoJu18,BaBePa19}. Let $\mu:=\Law(M_T)$. By \cite[Proposition~1.1]{GoJu18}, we have
\[
    \inf_{\pi\in\Pi(\mu,\nu)}
    \int_{\RR}\left|x-\int_{\RR}y\,\pi^x(dy)\right|^2\mu(dx)
    =
    \min_{\eta\preceq_c\nu}\Wcal_2(\mu,\eta).
\]
The unique minimizer $\eta^*$ is the $W_2$-projection of $\mu$ onto the set of laws dominated by $\nu$ in convex order. For any optimal $\pi$, its barycentric map $T(x):=\int y\,\pi^x(dy)$ is non-decreasing and $1$-Lipschitz and satisfies $T_\#\mu=\eta^*$; see \cite[Theorem~1.2]{GoJu18}. Thus, the optimal transport factors into the deterministic contraction $T$ followed by a martingale transport from $\eta^*$ to $\nu$. By \cite[Theorem~2.1]{GoJu18}, $\eta^*=\nu$ if and only if a non-decreasing $1$-Lipschitz map transports $\mu$ to $\nu$, which is precisely the feasibility condition in Theorem~\ref{mainThm}.

Equivalently, the distributions that can be obtained directly from $\mu$ through a non-decreasing $1$-Lipschitz map admit the following quantile characterization:
\begin{equation}\label{eq:attainable-quantiles}
  0\leq Q_\nu(v)-Q_\nu(u)
  \leq Q_\mu(v)-Q_\mu(u),
  \qquad 0<u<v<1.
\end{equation}
Thus, an attainable target distribution may contract the quantile increments of $\mu$, but cannot enlarge them. In particular, if $Q_\mu$ is constant on an interval corresponding to an atom of $\mu$, then $Q_\nu$ must also be constant on that interval.

\begin{proof}[Proof of Theorem~\ref{mainThm}]
Since both $S$ and $S^R$ are pure-jump processes, we have
	\begin{equation*}
		\sum_{t \in [0,T]}\EE[|\Delta S_t^R - \Delta S_t|^2] = \EE[\langle S - S^R \rangle_T] = \EE[|M_T-M^R_T|^2].
	\end{equation*}
Therefore, minimizing the functional in \eqref{mainPb} is equivalent to solving the quadratic optimal transport problem \eqref{MK} with initial distribution $\Law(M_T)$, target distribution $\hat \nu := \nu \ast \delta_{-\text{mean}(\nu)}$, and cost given by the squared Euclidean distance.

By Brenier's Theorem, this problem admits a unique solution, which is the coupling $(M_T, \hat F(M_T))$, where $\hat F$ is the optimal transport map from $\Law(M_T)$ to $\hat \nu$. Thus, it holds $\hat M_T^R = \hat F(M_T)$. Since $\hat M_T^R$ is an $\Fcal^M$-martingale and $M$ is Markovian, we have
\begin{equation*}
	M_t^R = \EE[\hat F(M_T)|\Fcal^M_t] = \EE[\hat F(M_T)|M_t] = F_t(M_t), \quad t \in [0, T],
\end{equation*}
where $F_t = \Law(M_T-M_t) \star \hat F$ is non-decreasing and $1$-Lipschitz, for any $t \in [0,T]$.
These two properties imply that $\hat S^R$ is non-decreasing and satisfies $0 \leq \Delta S^R_t \leq \Delta S_t$ a.s., for any $t \in [0, T]$.
\end{proof}

\begin{remark}[Examples of attainable target laws]
    Let $\mu = \Law(M_T)$. For $a \in [0,1]$ and $b \in \RR$, the affine contraction $F(x)=ax+b$ is non-decreasing and $1$-Lipschitz. Hence every target law of the form
    \[
        \nu=(a\,\id+b)_\#\mu
    \]
    is attainable.
    Since $M_T$ is centered, the construction in Theorem \ref{mainThm} gives $\hat M_t^R = aM_t$ and therefore induces a quota-share treat with retention rate $a$.
    Likewise, the maps $F(x)=x\wedge d$ and $F(x)=x\vee d$ are non-decreasing and $1$-Lipschitz and therefore produce attainable truncated target laws.
\end{remark}

\begin{remark}
If $U$ is the Cramér--Lundberg model \eqref{CL-model} and $\Law(\xi_i)$ is a non-atomic measure, for instance $\Law(\xi_i)\ll \lambda$, then the existence of a non-decreasing transport map $F$ as required in Theorem \ref{mainThm} is equivalent to $\nu$ being of the form
\begin{equation}
\label{CL-cond1}
	\nu = e^{-\lambda T}\delta_{x_0} + (1-e^{-\lambda T}) \rho, \quad \rho \in \Pcal_2(\RR),\; \supp(\rho) \subseteq (-\infty, x_0].
\end{equation}
Under this condition, the optimal transport map $F$ is given by
\begin{equation*}
	F(x) := \begin{cases}
		Q_\rho(F_{M_T|M_T \leq \overline \xi \lambda T}(x)) & \text{if } x \leq \overline \xi \lambda T, \\
		x_0 & \text{otherwise}.
	\end{cases}
\end{equation*}
Importantly, the measure $\rho$ represents the conditional distribution of $\nu$ given that at least one claim occurred, i.e., given the event $\{\omega \in \Omega: M_T(\omega) \leq \overline \xi \lambda T\}$. Since it is natural to assume that the surplus process evolves deterministically in the absence of claims, the condition \eqref{CL-cond1} on $\nu$ is not overly restrictive. 

Note that, if $S$ is modeled via a Gamma process, the existence of a non-decreasing transport map $F$ as required in Theorem \ref{mainThm} is guaranteed for any prescribed $\nu \in \Pcal_2(\RR)$, as a consequence of Brenier's Theorem. However, this map is not necessarily $1$-Lipschitz since this property depends on the choice of $\nu$.
\end{remark}

\begin{remark}[Stability of the optimizer]
	A common approach in Ruin Theory is to define a sequence of Cramér--Lundberg models $(U^{(n)})_{n \in \NN}$ with increasing intensity and scaled claim distributions, such that they weakly converge to a drifted Brownian motion $\tilde B$; see \cite[p. 226]{Sc07}. This approach is particularly convenient because it simplifies the model, making it more tractable from an analytical perspective.
	Let $M^{(n)}$ denote the martingale part of the process $U^{(n)}$. We can then conclude that the CDF $F_{M^{(n)}_t}$ of $M^{(n)}_t$ converges pointwise to the CDF of a centered Gaussian distribution.
	Assuming that the optimal transport map $\hat F^{(n)}$ from $M_T^{(n)}$ to $\hat \nu$ exists and is $1$-Lipschitz, it follows that the sequence of surplus processes resulting from the reinsurance strategy in Theorem \ref{mainThm} converges in distribution to the classical Bass martingale.
\end{remark}

In the next section, we use the prescribed-law construction developed here to solve the constrained reinsurance problem in which the terminal distribution is allowed to vary within a class determined by moment or risk-based requirements.

\section{Main Problem and Characterization of Optimal Reinsurance Strategies}\label{mainContrib}

We now turn to the main reinsurance problem of the paper. In the previous section, the terminal law of the reinsured surplus was fixed in advance. That prescribed-law problem provides the optimal-transport building block for the more flexible setting considered here, where the terminal distribution is not fixed but is only required to satisfy prescribed constraints, such as moment or risk-based constraints.

Let $\Acal_c$ denote the set of admissible terminal distributions satisfying the imposed constraints. More precisely, we set
\[
	\Acal_c = \Acal \cap \{\text{constraints}\},
\]
where
\[
    \Acal = \left\{\nu \in \Pcal_2(\RR) : \exists\, F \in \mathrm{Lip}_1(\RR) \text{ non-decreasing such that } F(M_T)\sim \nu\right\}.
\]
By \eqref{eq:attainable-quantiles}, the class $\Acal$ admits an equivalent quantile description. Namely, $\nu \in \Acal$ if and only if its quantile function satisfies
\[
    0 \le Q_\nu(v) - Q_\nu(u) \le Q_{M_T}(v) - Q_{M_T}(u), \qquad 0 < u < v < 1.
\]
Since translating $\nu$ does not change its quantile increments, the same condition characterizes the centered terminal laws used below.

The constrained reinsurance problem is then
\begin{equation}
 \label{mainProblem2}
 	\inf_{\nu\in \Acal_c} \,\inf_{\substack{U^R \text{ is admissible} \\ U_T^R \sim \nu}} \sum_{t \in [0,T]} \EE \left[ |\Delta S_t^R - \Delta S_t|^2  \right].
\end{equation}
Thus, the insurer chooses, among all terminal laws satisfying the prescribed constraints, the one that can be attained with minimal $L^2$-norm of the ceded risk. The solution of the prescribed-law problem from the previous section allows us to reduce \eqref{mainProblem2} to an optimization problem over quantile functions.

Under the assumption that the prescribed constraints can be expressed in terms of the quantile function of the terminal distribution of the surplus process, the following theorem shows that \eqref{mainProblem2} can be efficiently approached numerically.

 \begin{theorem}[ Solution of the reinsurance problem \eqref{mainProblem2}]
 \label{mainThm2}
     Assume there exists an optimizer $\tilde Q^* \in \Qcal_c$ of 
 \begin{equation}
 \label{equi_Prob2}
 	\inf_{\tilde Q \in \Qcal_c} \int_0^1 |\tilde Q(u)-Q_{M_T}(u)|^2 du, 
  \end{equation}
  where $\Qcal_c = \Qcal \cap \{\text{constraints}\}$ and 
  \begin{align*}
      \Qcal = \Bigg \{ \tilde Q \in L^2(0,1) : & \int_0^1 \tilde Q(u)du = 0  \text{ and } \\ & 0 \leq \tilde Q(x)-\tilde Q(y) \leq Q_{M_T}(x)-Q_{M_T}(y), \text{ for all } 0<x<y<1    \Bigg \}.
  \end{align*}
  Then the martingale part of any optimizer of \eqref{mainProblem2} is given by 
  \begin{equation*}
      \hat M_t^R = \EE[\hat Q(F_{M_T}(M_T))|M_t].
  \end{equation*}
  Additionally, this martingale is unique if and only if the optimizer of \eqref{equi_Prob2} is unique.
 \end{theorem}
 
\begin{remark}
If the set $\Qcal_c$ is convex and closed, \eqref{mainProblem2} becomes a convex optimization problem that can be solved efficiently. Moreover, if $\mathcal{Q}_c$ is nonempty, a solution to \eqref{mainProblem2} exists and is unique.
\end{remark}

 \begin{proof}
     It follows from Theorem \ref{mainThm} that for any $\nu \in \Acal_c$ and any admissible reinsurance strategy $U_t^R = u_0^R + g^R(t) + A_t^R- S_t^R$ with $U_T^R \sim \nu$, we have
 \begin{equation*}
 	\sum_{t \in [0,T]}\EE \left[ |\Delta S_t^R - \Delta S_t|^2  \right] = \EE[|M_T^R-M_T|^2] \geq \EE[|\hat F(M_T)-M_T|^2],
 \end{equation*}
 where $M_T^R \sim \hat \nu = \delta_{-\text{mean}(\nu)} \ast \nu$ and $\hat F \in \text{Lip}_1(\RR)$ is a non-decreasing function such that $\tilde F(M_T) \sim \tilde \nu$.
 
 If the law of $M_T$ is absolutely continuous, then,  by Brenier's Theorem,  $\hat F = Q_{\hat \nu}\circ F_{M_T}$. Hence,  we have
 \begin{equation*}
     \EE[|\hat F(M_T)-M_T|^2] = \EE[|Q_{\hat \nu}(U)-Q_{M_T}(U)|^2],
 \end{equation*}
 where $U\sim\text{Unif}_{[0,1]}$. In particular, $\hat F \in \text{Lip}_1(\RR)$ if and only if \[
 0 \leq Q_\nu(x)-Q_\nu(y) \leq Q_{M_T}(x)-Q_{M_T}(y),\quad \text{for all\ $0 < x < y < 1$}.
 \]
 Therefore, $U$ is optimal for \eqref{mainProblem2} if and only if the quantile function of $\hat \nu$, $Q_{\hat \nu}$ is an optimizer of \eqref{equi_Prob2}.
 
 On the other hand, if the law of $M_T$ is not absolutely continuous, then $\hat\nu$ is also not absolutely continuous. This follows from the existence of a function $F$ such that $F(M_T) \sim \nu$.  Furthermore, since $F$ is non-decreasing, each atom of $M_T$ is mapped to an atom of $\hat \nu$ under a monotone transformation, and the size of each atom in $\hat \nu$ is at least as large as that of the corresponding atom in $M_T$. Therefore, we still have $\hat F = Q_{\tilde \nu}\circ F_{M_T}$ and $0 \leq Q_{\hat \nu}(x)-Q_{\hat \nu}(y) \leq Q_{M_T}(x)-Q_{M_T}(y)$, for any $0 < x < y < 1$.
 Thus, we can identify any element of $\Acal_c$ with the quantile of its corresponding centered distribution, and any optimizer of \eqref{equi_Prob2} determines the martingale part of an optimal solution to \eqref{mainProblem2} and vice versa.
 \end{proof}

 \begin{remark}
    The optimizer of \eqref{equi_Prob2} does not uniquely determine the initial capital $u_0^R$ of the optimal reinsurance strategy, leaving the reinsurer free, in principle, to choose it arbitrarily.
 \end{remark}

We conclude this section with a few illustrative numerical applications of the optimization framework developed above. We focus on constraints imposed at the terminal time, although analogous constraints can also be imposed on marginals at intermediate times.
Notably, the numerical experiments below and the corresponding figures show that the monotone transport map $\widehat F$ associated with a solution $\widehat{\nu} \in \Acal_c$ of \eqref{mainProblem2}, which maps the law of the martingale part of the surplus process before reinsurance to that after reinsurance, often exhibits a structure related to classical quota-share and stop-loss treaties. In many cases, the domain of $\widehat F$ can be divided into intervals on which $\widehat F$ has the same form as the maps associated with classical quota-share or stop-loss treaties (see Figures \ref{fig:variance}, \ref{fig:VaR}, and \ref{fig:ES}). In other cases, this structure is observed only on certain intervals (see Figures \ref{fig:skewness}, \ref{fig:kurtosis}, and \ref{fig:all}). However, we stress that this observation is specific to the terminal-time problem, since the conditional-expectation construction leading to the optimal dynamic strategy need not preserve this structure at intermediate times, except in the variance case (see Proposition \ref{prop:variance_optimizer}).

\begin{remark}
	Let $\hat \nu \in \Pcal_2(\RR)$ be a probability measure with mean zero. Consider the process $(\hat M_t^R)_{t \in [0, T]}$, defined in Theorem \ref{mainThm}, such that $\hat M_T^R \sim \hat \nu$. Then the quantile function of $\hat M_t^R$ is given by
	\[
	Q_{M_t^R}^{Q_{\hat \nu}}(x) = \int_\RR Q_{\hat \nu}(F_{M_T}(Q_{M_t}(x)+y)) \Law(M_T-M_t)(dy),
	\]
where the superscript emphasizes the dependence on the quantile function $Q_{\hat{\nu}}$.
Importantly, the map $Q \mapsto Q_{M_t^R}^{Q}$ is linear. Consequently, if the optimization problem \eqref{equi_Prob2} is convex under a given set of constraints applied to the terminal marginal, it remains convex when the same constraints are applied to any intermediate marginal of the process $(M_t^R)_{t \in [0, T]}$.
\end{remark}

 \subsection{Variance constraint}
 One natural application arises when the goal is to limit the risk exposure by controlling the variance of the surplus process at terminal time. More specifically, we consider the problem to reduce the variance of $U_T^R$ under a given threshold $0<k<{\rm Var}(U_T)$, while simultaneously minimizing the $L^2$-norm of the ceded risk. This corresponds to problem \eqref{mainProblem2} with $\Acal_c = \Acal \cap \{\nu \in \Pcal(\RR): \text{Var}(\nu) \leq k\}$. Since every element in $\Qcal$ has mean zero and ${\rm Var}(U_T^R) =  {\rm Var}(M_T^R)$ for any admissible reinsurance strategy, the constraint translates into
 \begin{equation*}
     \Qcal_c = \Qcal \cap \left\{\tilde Q \in L^2(0,1) : \int_0^1 \tilde Q^2(u)du \leq k \right\}.
 \end{equation*}
Therefore, problem \eqref{equi_Prob2} becomes a convex optimization problem. In particular, its optimizer admits an explicit representation.

 \begin{proposition}
 \label{prop:variance_optimizer}
 	The unique optimizer of \eqref{equi_Prob2} under variance constraint is given by
 	\begin{equation*}
 		\hat Q = \sqrt \frac{k}{{\rm Var}(M_T)} Q_{M_T}.
 	\end{equation*}
In particular, among all admissible strategies that reduce variance below the threshold $k$, the quota-share contract minimizes the $L^2$-norm of the ceded risk. 
\end{proposition}
\begin{proof}
	Let $\tilde Q \in \Qcal_c$. By the triangular inequality, we have
	\begin{align*}
		\int_0^1 |\tilde Q(u)-Q_{M_T}(u)|^2 du & \geq \left( \sqrt{\int_0^1 \tilde Q^2(u)du} - \sqrt{\int_0^1 Q^2_{M_T}(u)du}\right)^2 \geq \left(\sqrt{k} - \sqrt{{\rm Var}(M_T)}\right)^2.
	\end{align*}
	This bound is attained by $\hat Q$, and uniqueness follows directly from the fact that quantile functions are non-decreasing.
\end{proof}

\begin{figure}
     \centering
     \begin{subfigure}[t]{0.49\textwidth}
         \centering
         \includegraphics[width=\textwidth]{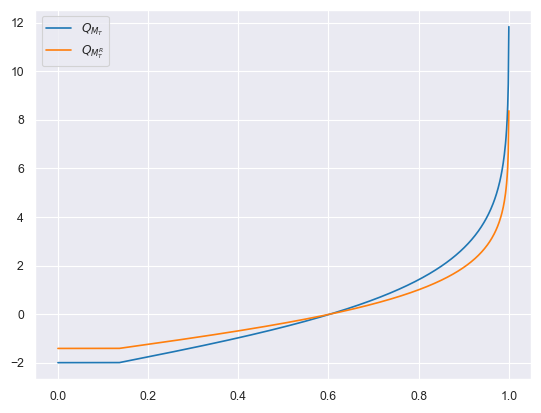}
         \caption{The optimization problem.}
     \end{subfigure}
     \hfill
     \begin{subfigure}[t]{0.49\textwidth}
         \centering
         \includegraphics[width=\textwidth]{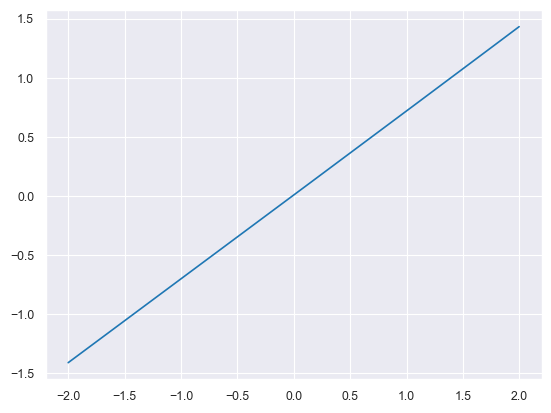}
         \caption{The transport map $\hat F$ from $M_T$ to the optimal $M_T^R$.}
     \end{subfigure}
    \captionsetup{singlelinecheck=false}
	\caption{Optimization under variance constraint: $\text{Var}(M_T^R) \leq \text{Var}(M_T)/2$. $M$ is a compensated compound Poisson process with intensity $1$ and the claims size is exponentially distributed with parameter $1$}.
    \label{fig:variance}
\end{figure}

Figure \ref{fig:variance} provides numerical confirmation that the optimal reinsurance strategy for reducing variance, while minimizing the $L^2$-norm of the ceded risk, is given by the quota-share contract.

 \subsection{Value-at-Risk and Expected Shortfall constraints}
In this section, we assume that the reinsured surplus process at terminal time must satisfy either a Value-at-Risk constraint $Q_{U_T^R}(p) \geq k$ or an Expected Shortfall constraint $\frac{1}{p} \int_0^p Q_{U_T^R}(u) du \geq k$, where $k>0$ is a prescribed threshold.
 Since the admissible reinsurance strategies are deterministically-drifted martingales, we have
 \begin{equation*}
 	Q_{U_T^R}(p) = Q_{M_T^R}(p) + u_0^R + g^R(T)  \geq k \iff Q_{M_T^R}(p) - h_{Q_{M_T^R}}(T) \geq k - u_0^R - g(T),
 \end{equation*} 
 where we use the subscript $Q_{M_T^R}$ on $h$  to stress that the reinsurance premium may depend on the quantile function of $M_T^R$, even though the surcharge function $h$ is deterministic. Therefore, we have 
 \begin{align*}
     \Qcal_c = \Qcal \cap
     \left\{\tilde Q \in L^2(0,1) :  \tilde Q(p) - h_{\tilde Q}(T) \geq k - u_0^R - g(T) \right\}.
 \end{align*}
 In particular, $\Qcal_c$ is convex if and only if $h_{\tilde Q}$ is a convex functional of $\tilde Q$. This condition is satisfied, for example, when the deterministic surcharge is defined by 
  \begin{equation*}
 	h_{Q_{M_T^R}}(t) = \frac{\alpha t}{T}\sqrt{\sum_{s \in [0,T]}\EE \left[ |\Delta S_t^R - \Delta S_t|^2  \right]} = \frac{\alpha t}{T} \|Q_{M_T^R}-Q_{M_T}\|_{L^2(0,1)},
 \end{equation*}
 for some $\alpha > 0$. Similarly, the Expected Shortfall constraint is equivalent to 
  \begin{equation*}
 	\frac{1}{p}\int_0^p Q_{M_T^R}(u) du - h_{Q_{M_T^R}}(T) \geq k - u_0^R - g(T)
 \end{equation*}
 and we have
  \begin{align*}
     \Qcal_c = \Qcal \cap
     \left\{\tilde Q \in L^2(0,1) :  \frac{1}{p}\int_0^p Q_{M_T^R}(u) du - h_{\tilde Q}(T) \geq k - u_0^R - g(T) \right\},
 \end{align*} 
 which is a convex set if we assume that $h_{\tilde Q}$ is a convex functional of $\tilde Q$.
 
\begin{figure}
     \centering
     \begin{subfigure}[t]{0.49\textwidth}
         \centering
         \includegraphics[width=\textwidth]{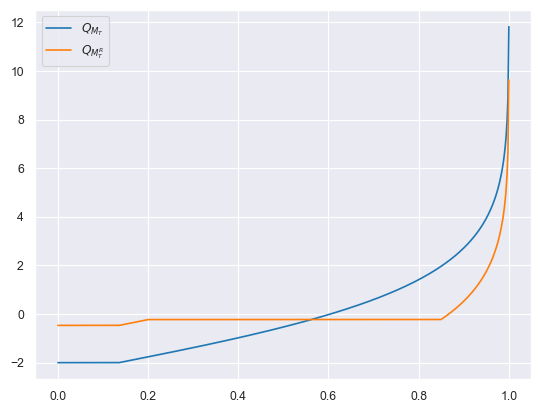}
        \captionsetup{justification=raggedright, singlelinecheck=false}
         \caption{The optimization problem.}
     \end{subfigure}
     \hfill
     \begin{subfigure}[t]{0.49\textwidth}
         \centering
         \includegraphics[width=\textwidth]{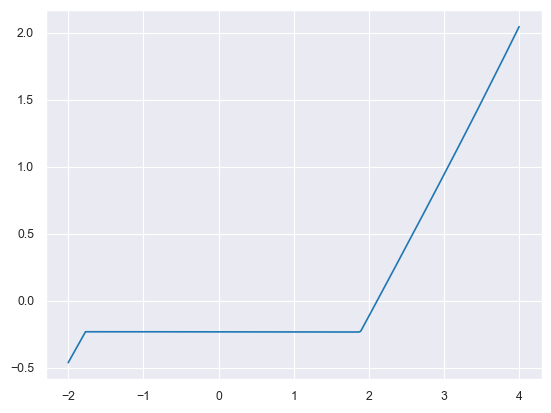}
        \captionsetup{justification=raggedright, singlelinecheck=false}
         \caption{The transport map $\hat F$ from $M_T$ to the optimal $M_T^R$.}
     \end{subfigure}
    \captionsetup{singlelinecheck=false}
     \caption{Optimization under Value-at-Risk constraint: $Q_{M_T^R}(0.2) - 0.05 \|M_T^R - M_T \|_{L^2} \geq -0.3$. $M$ is a compensated compound Poisson process with intensity $1$ and the claim size is exponentially distributed with parameter $1$.}
      \label{fig:VaR}
\end{figure}

\begin{figure}
     \centering
     \begin{subfigure}[t]{0.49\textwidth}
         \centering
         \includegraphics[width=\textwidth]{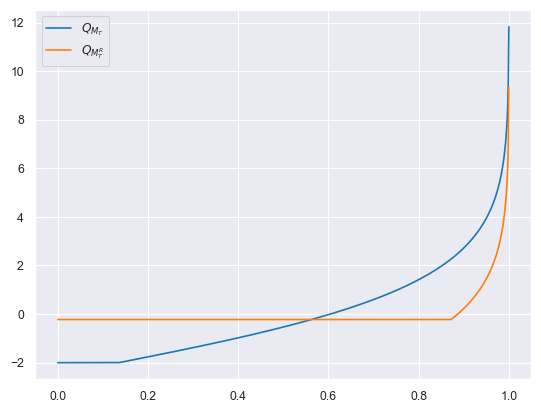}
        \captionsetup{justification=raggedright, singlelinecheck=false}
         \caption{The optimization problem.}
     \end{subfigure}
     \hfill
     \begin{subfigure}[t]{0.49\textwidth}
         \centering
         \includegraphics[width=\textwidth]{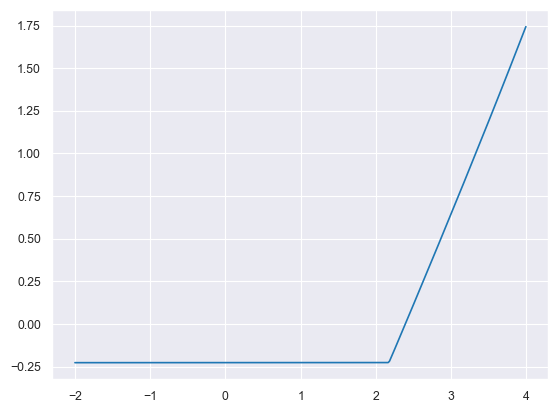}
        \captionsetup{justification=raggedright, singlelinecheck=false}
         \caption{The transport map $\hat F$ from $M_T$ to the optimal $M_T^R$.}
     \end{subfigure}
    \captionsetup{singlelinecheck=false}
     \caption{Optimization under Expected Shortfall constraint: $5\int_0^{0.2} Q_{M_T^R}(u) du - 0.05 \|M_T^R - M_T \|_{L^2} \geq -0.3$. $M$ is a compensated compound Poisson process with intensity $1$ and the claim size is exponentially distributed with parameter $1$.}
       \label{fig:ES}
\end{figure}

Figure \ref{fig:VaR} illustrates the solution to the optimization problem \eqref{equi_Prob2} with a Value-at-Risk constraint, while Figure \ref{fig:ES} presents the solution to the same optimization problem with an Expected Shortfall constraint. The optimal solution can vary depending on the specific constraint. In this case, the optimal reinsurance strategy at terminal time $T$ under the Value-at-Risk constraint combines quota-share and stop-loss reinsurance, while under the Expected Shortfall constraint, the optimal strategy at terminal time $T$ is a pure stop-loss contract.
  
 \subsection{Skewness and Kurtosis constraints}
 Finally, to control the risk profile of the reinsurance strategy at the terminal time, we may impose constraints on the skewness and kurtosis of the distribution of $U_T^R$. A positively skewed distribution, which has a longer right tail, is typically desirable, as it implies that extreme gains are more likely than extreme losses. In contrast, kurtosis quantifies the heaviness of a distribution's tails. High kurtosis is generally undesirable, as it indicates a higher probability of extreme outcomes. Imposing bounds on skewness and kurtosis can thus help mitigate tail risk and promote a more favorable risk-return profile for the reinsured surplus.
 
 The skewness of a random variable $X$ is  defined as
 \begin{equation*}
 	\text{Skew}[X] = \EE \left[ \left(\frac{X-\mu_X}{\sigma_X} \right)^3 \right] = \int_0^1 \left(\frac{Q_X(u)-\mu_X}{\sigma_X} \right)^3 du,
 \end{equation*}
 where $Q_X$ is the quantile function of $X$, $\mu_X = \int_0^1 Q_X(u) du$ is the mean of $X$ and $\sigma_X^2 = \int_0^1 (Q_X(u) - \mu_X)^2 du$ is the variance of $X$. However, using the above (standard) definition of skewness, the constraint $\text{Skew}[U_T^R] \geq k$, where $k>0$, is non-convex, making the optimization problem \eqref{equi_Prob2} potentially difficult to solve. To address this issue, a more practical approach might be to use a quantile-based measure of skewness. Groeneveld and Meeden (see \cite{GrMe84}) proposed a suitable alternative, defined as
 \begin{equation}
 \label{skewness}
 	\text{Skew}_u[X] = \frac{Q_X(1-u) + Q_X(u) - 2Q_X(1/2)}{Q_X(1-u) - Q_X(u)} \in [-1, 1],
 \end{equation}
 where $u$ is a chosen parameter in $(0, 1/2)$. This measure captures the skewness of the distribution based on quantiles. To obtain a more global measure of skewness, several extensions based on \eqref{skewness} have been introduced in the literature, such as 
 \begin{equation*}
 	\text{Skew}_{\text{sup}}[X] = \sup_{u \in (0, 1/2)}\text{Skew}_u[X] \quad \text{or} \quad \text{Skew}_{\text{int}}[X] = \frac{\int_0^1 Q_X(1-u) - Q_X(u) - 2Q_X(1/2) du}{\int_0^1 Q_X(1-u) - Q_X(u) du}.
 \end{equation*}
 Together with Pearson’s median skewness (also known as Pearson’s second skewness coefficient)
 \begin{equation*}
 	\text{Skew}_{\text{Pearson, 2}}[X] = \frac{3(\mu_X - Q_X(1/2))}{\sigma_X},
 \end{equation*}
these definitions makes the constraint ``skewness of $U_T^R$ greater of $k$" (and thus problem \eqref{equi_Prob2}) convex. Hence, although these definitions lead to different notions of skewness, any of them may be used, with the choice depending on the context. Notably, the measure $\text{Skew}_{\text{int}}[X]$ is closely related to $\text{Skew}_{\text{Pearson, 2}}[X]$, as shown by the following expression
 \begin{equation*}
 	\text{Skew}_{\text{int}}[X] = \frac{\mu_X - Q_X(1/2)}{\EE[|X - Q_X(1/2)|]}.
 \end{equation*} 
 
The kurtosis of a random variable $X$ is commonly defined as
 \begin{equation*}
 	\text{Kurt}[X] = \EE \left[ \left(\frac{X-\mu_X}{\sigma_X} \right)^4 \right] = \int_0^1 \left(\frac{Q_X(u)-\mu_X}{\sigma_X} \right)^4 du,
 \end{equation*}
 where $Q_X$ is the quantile function of $X$, $\mu_X = \int_0^1 Q_X(u) du$ is the mean of $X$ and $\sigma_X^2 = \int_0^1 (Q_X(u) - \mu_X)^2 du$ is the variance of $X$. This definition, however, results in a non-convex optimization problem when we impose the constraint $\text{Kurt}[U_T^R] \leq k$, $k>0$. As an alternative, we can use the quantile-based measure of kurtosis introduced by Ruppert in \cite{Ru87}, given by
 \begin{equation*}
 	\text{Kurt}_{u, v}[X] = \frac{Q_X(1-u)-Q_X(u)}{Q_X(1-v)-Q_X(v)},
 \end{equation*} 
 for $0 < u < v < 1/2$ chosen.
 
\begin{figure}
     \centering
     \begin{subfigure}[t]{0.47\textwidth}
         \centering
         \includegraphics[width=\textwidth]{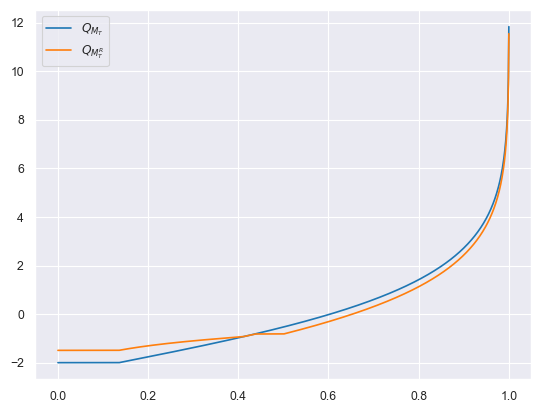}
         \caption{The optimization problem.}
     \end{subfigure}
     \hfill
     \begin{subfigure}[t]{0.47\textwidth}
         \centering
         \includegraphics[width=\textwidth]{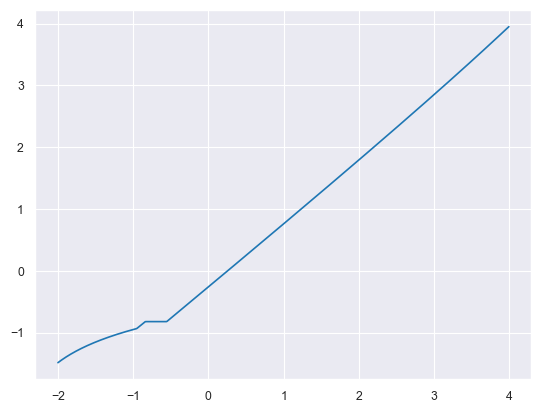}
         \caption{The transport map $\hat F$ from $M_T$ to the optimal $M_T^R$.}
     \end{subfigure}
     \caption{Optimization under Skewness constraint: $\text{Skew}_{\text{sup}}[Q_{M_T^R}] \geq 0.6$. $M$ is a compensated compound Poisson process with intensity $1$ and the claim size is exponentially distributed with parameter $1$.}
     \label{fig:skewness}
\end{figure}

\begin{figure}
     \centering
     \begin{subfigure}[t]{0.47\textwidth}
         \centering
         \includegraphics[width=\textwidth]{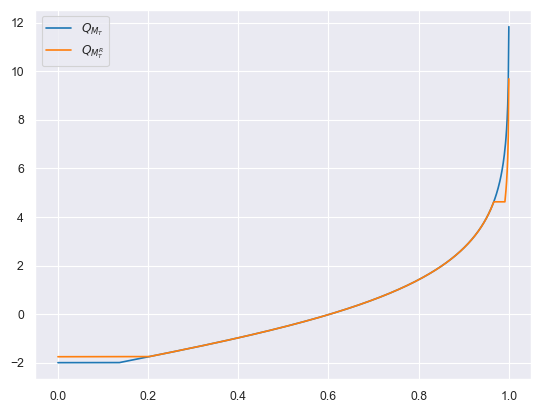}
         \caption{The optimization problem.}
     \end{subfigure}
     \hfill
     \begin{subfigure}[t]{0.47\textwidth}
         \centering
         \includegraphics[width=\textwidth]{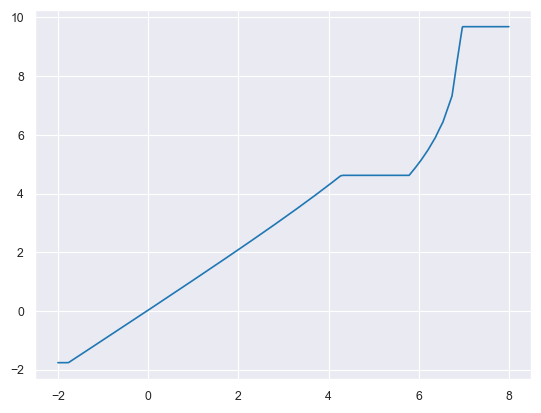}
         \caption{The transport map $\hat F$ from $M_T$ to the optimal $M_T^R$.}
     \end{subfigure}
     \caption{Optimization under Kurtosis constraint: $\text{Kurt}_{0.1, 0.25, v}[X] \leq 2.5$. $M$ is a compensated compound Poisson process with intensity $1$ and the claim size is exponentially distributed with parameter $1$.}
      \label{fig:kurtosis}
\end{figure}

\begin{figure}
     \centering
     \begin{subfigure}[t]{0.47\textwidth}
         \centering
         \includegraphics[width=\textwidth]{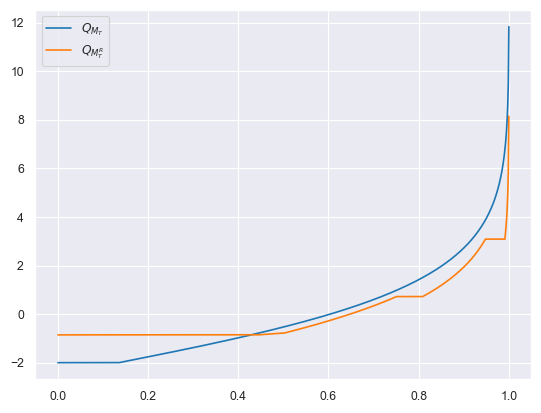}
         \caption{The optimization problem.}
     \end{subfigure}
     \hfill
     \begin{subfigure}[t]{0.47\textwidth}
         \centering
         \includegraphics[width=\textwidth]{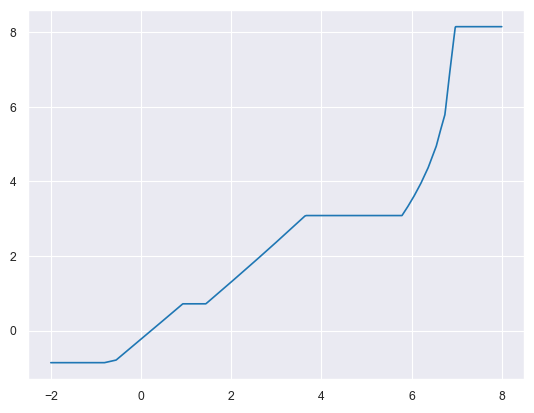}
         \caption{The transport map $\hat F$ from $M_T$ to the optimal $M_T^R$.}
     \end{subfigure}
    \caption{Optimization under multiple constraints: $\text{Var}(M_T^R) \leq \text{Var}(M_T)/2$, $5\int_0^{0.2} Q_{M_T^R}(u) du - 0.05 \|M_T^R - M_T \|_{L^2} \geq -0.9$, $\text{Skew}_{\text{sup}}[Q_{M_T^R}] \geq 0.6$, and $\text{Kurt}_{0.1, 0.25, v}[X] \leq 2.5$. $M$ is a compensated compound Poisson process with intensity $1$ and the claim size is exponentially distributed with parameter $1$.}
    \label{fig:all}
\end{figure}

 Figures \ref{fig:skewness}, \ref{fig:kurtosis}, and \ref{fig:all} illustrate the solutions to the optimization problem \eqref{equi_Prob2} under skewness, kurtosis, and multiple mixed constraints, respectively. In these cases, the optimal reinsurance strategy at terminal time $T$ is not a combination of a quota-share and stop-loss contract, as the transport map $\hat F$ appears to be piecewise quadratic. However, $\hat F$ in Figures \ref{fig:kurtosis} and \ref{fig:all} can be efficiently approximated by a combination of proportional and stop-loss contracts.
   
\printbibliography
\end{document}